\documentclass[conference]{IEEEtran}
\IEEEoverridecommandlockouts

\usepackage{amsthm}
\usepackage{booktabs}
\usepackage{algorithm}
\usepackage{algorithmic}
\usepackage{tikz}
\usepackage{tikz-qtree}
\usepackage{color, colortbl}
\usepackage{xspace}

\usepackage{cite}
\usepackage{amsmath,amssymb,amsfonts}
\usepackage{algorithmic}
\usepackage{graphicx}
\usepackage{textcomp}
\usepackage{xcolor}
\usepackage{verbatim}

\def\BibTeX{{\rm B\kern-.05em{\sc i\kern-.025em b}\kern-.08em
    T\kern-.1667em\lower.7ex\hbox{E}\kern-.125emX}}

\begin{document}

\title{\textsc{DirectDebug}: Automated Testing and\\ Debugging of Feature Models}


\author{\IEEEauthorblockN{Viet-Man Le}
\IEEEauthorblockA{\textit{Institute of Software Technology} \\
\textit{Graz University of Technology}\\
Graz, Austria \\
vietman.le@ist.tugraz.at}
\and
\IEEEauthorblockN{Alexander Felfernig}
\IEEEauthorblockA{\textit{Institute of Software Technology} \\
\textit{Graz University of Technology}\\
Graz, Austria \\
alexander.felfernig@tugraz.at}
\and
\IEEEauthorblockN{Mathias Uta}
\IEEEauthorblockA{\textit{Siemens } \\
\textit{Energy AG}\\
Erlangen, Germany \\
mathias.uta@siemens.com}
\and

  \hfill\mbox{}\par
  \mbox{}\hfill
\IEEEauthorblockN{David Benavides}
\IEEEauthorblockA{\textit{Computer Languages and Systems} \\
\textit{University of Sevilla}\\
Seville, Spain \\
benavides@us.es}
\and
\IEEEauthorblockN{Jose Galindo}
\IEEEauthorblockA{\textit{Computer Languages and Systems} \\
\textit{University of Sevilla}\\
Seville, Spain \\
jagalindo@us.es}
\and
\IEEEauthorblockN{Thi Ngoc Trang Tran}
\IEEEauthorblockA{\textit{Institute of Software Technology} \\
\textit{Graz University of Technology}\\
Graz, Austria \\
ttrang@ist.tugraz.at}
}

\maketitle

\begin{abstract}
Variability models (e.g., feature models) are a common way for the representation of variabilities and commonalities of software artifacts. Such models can be translated to a logical representation and thus allow different operations for quality assurance and other types of model property analysis. Specifically, complex and often large-scale feature models can become faulty, i.e., do not represent the expected variability properties of the underlying software artifact. In this paper, we introduce \textsc{DirectDebug} which is a direct diagnosis approach to the \emph{automated testing and debugging of variability models}. The algorithm helps software engineers by supporting an automated identification of faulty constraints responsible for an unintended behavior of a variability model. This approach can significantly decrease development and maintenance efforts for such models.
\end{abstract}


\section{Introduction}\label{introduction}

Feature models support the representation of variability and commonality properties of software artifacts \cite{BeSeRu2010,Kang1990}. Applications thereof support users in deciding about which features should be included in a specific software instance. These models can be differentiated with regard to the used knowledge representation. So-called \emph{basic feature models} \cite{Kang1990} support the representation of  hierarchies including cross-tree constraints such as \emph{excludes} and \emph{requires} relationships. \emph{Cardinality-based feature models} \cite{Czarnecki2005}  extend basic ones with cardinalities ($>1$) of feature relationships. Finally, \emph{extended feature models} \cite{Ba2005} support the description of features with attributes.

The creation and evolution of feature models can be error-prone where \emph{cognitive overloads} or \emph{missing domain knowledge} are major reasons for models that do not reflect the intended variability  properties \cite{Benavidesetal2013,Trinidad2008,ZellerIEEEComputer2001}. Consequently, feature model development has to be pro-actively supported by intelligent debugging mechanisms that support the automated detection of faulty constraints responsible for the unexpected behavior of a feature model knowledge base. 

The remainder of this paper is organized as follows. After a discussion of related work (Section \ref{relatedwork}), we introduce an example of a feature model (Section \ref{featuremodels}). In this context, we provide a formalization of feature models as constraint satisfaction problems (CSPs). This formalization is used as a basis for a discussion of concepts supporting the automated testing (Section \ref{testing}) and debugging (Section \ref{automatedtestingdebugging}) of feature models. In Section \ref{evaluation}, we report initial results of a performance analysis of our approach. The paper is concluded with  Section \ref{conclusions}. 

\section{Related Work}\label{relatedwork}

The state-of-the-art in feature model analysis and related tasks can be summarized as follows.

\textbf{Feature Model Analysis Operations}. Analysis operations help to assure well-formedness properties in feature models. For example, it should be possible that each feature of a model can be included in at least one  configuration, i.e., there should not exist a feature which is inactive in every possible configuration. For a detailed discussion of analysis operations for feature models we refer to Benavides et al. \cite{BeSeRu2010}.

\textbf{Conflict Detection}. A conflict set (conflict) can be regarded as a \emph{subset that induces an inconsistency}. For example, Zeller et al. \cite{Zeller2002}  propose the \textsc{Delta Debugging} algorithm which supports the determination of relevant subsets in test cases responsible for the faulty behavior of a software component. Following a similar objective, Junker \cite{Junker2004} introduces the \textsc{QuickXPlain} algorithm for the identification of  subsets of constraints in a knowledge base responsible for an inconsistency (no solution can be found). Conflict sets are the basis for follow-up diagnosis operations that help to resolve these conflicts. More precisely, a diagnosis (hitting set \cite{Felfernig2012,Reiter1987})  entails a  set of elements of a knowledge base that have to be adapted or deleted to be able to resolve all conflicts, i.e., to restore consistency in the knowledge base. In contrast to conflict detection \cite{Junker2004,Zeller2002}, the approach presented in this paper focuses on diagnosis, i.e., \emph{conflict resolution}.

\textbf{Diagnosis of Inconsistent Models}. A diagnosis can be regarded as a \emph{deletion subset that helps to restore consistency}. An approach to the identification of diagnoses for inconsistent constraint sets is presented in Bakker et al. \cite{Bakker93}. In this line of research, Trinidad et al. \cite{Trinidad2008} show how to determine such diagnoses in the context of inconsistent feature models. In contrast to the work of  Bakker et al. \cite{Bakker93} and Trinidad et al. \cite{Trinidad2008}, the approach presented in this paper focuses on scenarios where test cases are used to induce an inconsistency in a knowledge base. Our approach is based on the idea of Felfernig et al. \cite{FeFrJaSt2004} who introduce a model-based diagnosis approach \cite{Reiter1987} to resolve conflicts in knowledge bases induced by test cases. Compared to  Felfernig et al. \cite{FeFrJaSt2004}, our approach is based on \emph{direct diagnosis} (no conflict detection needed) which allows for an efficient determination of diagnoses \cite{Felfernig2012}.

\textbf{Diagnosis for Reconfiguration}.  A reconfiguration can be regarded as a \emph{set of adaptations of feature settings in a changed configuration  that are needed to restore consistency between the configuration and the corresponding feature model} \cite{Felfernig2018reconf}. Using a constraint-based representation \cite{Tsang1993} of a feature model, White et al. \cite{White2010} show how to apply the concepts of model-based diagnosis \cite{Reiter1987} to determine minimal sets of feature settings in existing configurations that need to be adapted in order to restore consistency with the feature model. In this context, feature models are assumed to be consistent. The work presented in this paper generalizes the concepts of Trinidad et al. \cite{Trinidad2008} and White et al. \cite{White2010} by allowing to take into account \emph{a set of test cases at the same time}, i.e., a diagnosis represents an adaptation proposal that makes all of the given test cases consistent with the knowledge base.

\textbf{Direct Diagnosis}. The idea of \emph{direct diagnosis} \cite{Felfernig2012} is to significantly improve the performance of hitting set based diagnosis approaches \cite{Reiter1987} by supporting the calculation of diagnoses without the need of predetermining conflict sets. In this paper, we show how to support the automated testing and debugging of feature models using direct diagnosis \cite{Felfernig2012}.
 
The \emph{contributions} of this paper are threefold.  \emph{First}, we show how to extend  direct diagnosis \cite{Felfernig2012} to support the automated testing and debugging of feature models. We integrate testing and diagnosis in a unified manner where test cases are considered as a central element of a diagnosis process. \emph{Second}, we show how different types of test cases can be integrated into automated debugging processes. \emph{Third}, we report initial results of a performance analysis of our diagnosis approach.

\section{Feature Model Semantics}\label{featuremodels}
Feature models are used to represent software variability properties by specifying features and their relationships in a hierarchical fashion \cite{Kang1990}. Features are arranged in a hierarchical fashion with one specific root feature $f_r$ which has to be included in every configuration \cite{BeSeRu2010}. In a feature model, features are represented as nodes and relationships between features as corresponding edges. For an overview of  approaches to the representation of feature models, we refer to Batory \cite{Ba2005}.

\textbf{Feature Model Semantics}. For the discussions in this paper, we follow the feature model representation of Benavides et al. \cite{BeSeRu2010} which includes four types of relationships (the hierarchical constraints  \emph{mandatory}, \emph{optional}, \emph{alternative}, \emph{or}) and two types of cross tree constraints (\emph{requires} and \emph{excludes}). Feature models are variability models which can be formalized as a \emph{Constraint Satisfaction Problem} (CSP) \cite{Tsang1993}. Each  feature $f_i$ is related to the binary domain $\{(t)rue,(f)alse\}$. The mentioned relationships and cross-tree constraints are represented as constraints on the CSP level. 

\begin{figure}[ht]
			\centering
			\fbox{
				\includegraphics[width=0.35 \textwidth]{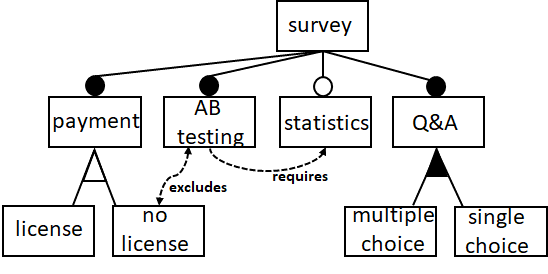}
			}
			\caption{An example of a (presumably faulty) \emph{survey software} feature model.}
			\label{Figure1}
		\end{figure}

\textbf{Example Feature Model}. Figure \ref{Figure1} depicts an example of a \emph{presumably faulty feature model} (for details, see Section \ref{testing}) from the domain of \emph{software services supporting the creation and management of surveys}. For example, the feature \emph{ABtesting} indicates whether a user wants to use AB testing functionalities when analyzing the results of a user study completed on the basis of questionnaires. Furthermore, the feature \emph{payment} indicates the preferred payment mode where \emph{license} represents a yearly payment and \emph{nolicense} indicates a free license with an associated limited set of enabled features.


\textbf{Constraint Types}. The following semantics of feature model constraints is based on Benavides et al. \cite{BeSeRu2010}.

	\textbf{Mandatory}: feature $f_b$ is denoted as \emph{mandatory} if it is in a mandatory relationship with another feature $f_a$. On the logical level, a mandatory relationship is defined in terms of an equivalence $f_a$ $\leftrightarrow$ $f_b$. In Figure \ref{Figure1}, the feature \emph{Q\&A} is mandatory, i.e., it has to be part of every \emph{survey} configuration. The same holds for \emph{payment} and \emph{ABtesting} where the latter should be considered faulty since, for example, it makes \emph{nolicense} a dead feature and \emph{statistics} a false optional \cite{BeSeRu2010}.
	
	\textbf{Optional}: if a feature $f_b$ is denoted as optional, this means that it may or may not be included in the case that feature $f_a$  is included. On the logical level, this property is formulated as implication: $f_b$ $\rightarrow$ $f_a$. In Figure \ref{Figure1}, \emph{statistics} is an optional feature connected to \emph{survey}. 

	\textbf{Alternative}: exactly one feature $f_i$ out of $\{f_1, .., f_k\}$ has to be selected if feature $f_a$ has been selected. On the logical level, \emph{alternative} relationships can be formalized as follows: $f_1=t \leftrightarrow (f_2=f \land .. \land f_k=f \land f_a=t) \land .. \land f_k=t \leftrightarrow (f_1=f \land .. \land f_{k-1}=f \land f_a=t)$.  An example thereof is \emph{payment} with the subfeatures  \emph{license} and \emph{nolicense}.
	
	\textbf{Or}: at least one feature $f_i$ out of a feature set $\{f_1, .., f_k\}$ has to be selected if feature $f_a$ has been selected. Relationships of type \emph{or}  can be formalized as follows: $f_a$ $\leftrightarrow$ $f_1=t \lor f_2=t \lor .. \lor f_k = t$.  An example of a feature $f_a$ is \emph{Q\&A}, the  subfeatures are \emph{multiplechoice} and \emph{singlechoice}.

	\textbf{Requires}: feature $f_b$ must be included in a configuration if feature $f_a$ is included. On the logical level, \emph{requires} relationships can be defined as $f_a$ $\rightarrow$ $f_b$. An example of a \emph{requires} relationship is \emph{ABtesting} $\rightarrow$ \emph{statistics}.

	\textbf{Excludes}: $f_a$ and $f_b$ must not be combined ($f_a$ excludes feature $f_b$ and vice versa). On the logical level, \emph{excludes} relationships can be defined as $\neg(f_a \land f_b)$.  An example of an \emph{excludes} relationship is: $\neg(ABtesting \land nolicense)$.
	
	\textbf{Feature Models and Configuration Tasks}. The task of finding a solution for a constraint satisfaction problem representing a feature model can be interpreted as a configuration task (see Definition 1).

\textbf{Definition 1 (Configuration Task)}. A configuration task $(F,D,C)$ is defined by a feature set $F=\{f_1, f_2, ..., f_n\}$ and a set of feature domains $D=\{dom(f_1),$  $dom(f_2),$ $..., dom(f_n)\}$ ($dom(f_i)=\{(t)rue, (f)alse\}$). Furthermore, $C=CR \cup CF$ represents constraints restricting the possible solutions for a configuration task where $CR = \{c_1, c_2, ..., c_k\}$ is a set of user  requirements and $CF=\{c_{k+1}, c_{k+2}, ..., c_{m}\}$  a set of feature model constraints.

In Definition 1, $CR$ is an additional set of constraints which specifies which features should be included in a configuration.

	Based on Definition 1, we introduce the concept of a \emph{configuration} (solution) for a configuration task (Definition 2). 

	\textbf{Definition 2 (Configuration).} A feature model configuration for a given feature model configuration task is an assignment $A$ of all feature variables $f_i \in F$. $A$ is consistent if $A$ does not violate any constraint in $CR \cup CF$.
	
	\textbf{CSP Representation of a Feature Model}. A CSP-based representation of a feature model configuration task $(F,D,C=CR \cup CF)$ that can be generated from the  model shown in Figure 1 is the following. In this context, $c_0: survey = t$ is regarded as \emph{root constraint} that is used to avoid the derivation of (irrelevant) empty configurations.

	\begin{itemize} \small
			\item{$F=\{survey$, $payment$, $license$, $nolicense$, $ABtesting$, $statistics$, $Q\&A$, $multiplechoice$, $singlechoice\}$}
			\item{$D=\{dom(survey)$ = $\{t$, $f\}$, $dom(payment)$ = $\{t$, $f\}$, .., $ dom(singlechoice)$ $ = $ $\{t$, $f\}\}$}
			\item{$CF=\{c_0: survey=t$, $c_1:survey$ $\leftrightarrow payment$, $c_2:$ $survey$ $\leftrightarrow ABtesting$,  $c_3$: $statistics \rightarrow survey$, $c_4$: $survey \leftrightarrow Q\&A$, $c_5$: $Q\&A \leftrightarrow multiplechoice \lor singlechoice$, $c_{6}$: $(license \leftrightarrow \neg nolicense \land payment)$ $ \land $ $ (nolicense \leftrightarrow \neg license \land payment)\}$, $c_{7}$: $\neg(ABtesting \land nolicense)$,  $c_{8}$: $ABtesting \rightarrow statistics\}$}
			\item{$CR=\{c_9: license = t, c_{10}: ABtesting = t\}$}
	\end{itemize}	

    A consistent configuration that can be generated from our example feature model configuration task is the following:

    \begin{itemize} \small
			\item \{$survey=t$, $payment=t$, $license=t$, $nolicense=f$, $ABtesting=t$, $statistics=t$, $Q\&A= t$, $multiplechoice=t$, $singlechoice=t$\}
	\end{itemize}	

Due to faulty constraints in a feature model, in some situations the constraint solver (configurator) does not determine the intended solution(s). We now show how to identify a minimal set of constraints in a feature model that need to be adapted (or deleted) to restore consistency with regard to a predefined set of test cases which specify the intended behavior of a feature model. In other words, we are interested in constraints responsible for the faulty behavior of a knowledge base generated from a feature model. In the following, we will show how to automatically determine such constraint sets on the basis of the concepts of \emph{direct diagnosis} \cite{Felfernig2012}.

\section{Testing Feature Models}\label{testing}

In Figure 1, \emph{ABtesting} is mandatory, however, this triggers a situation where each configuration has to include \emph{statistics} and it is not possible to generate a configuration with \emph{nolicense} payment. Reasons for faulty feature models can range from \emph{misinterpretations in domain knowledge communication}, \emph{modeling errors}, to \emph{outdated parts of a knowledge base}.
		
\textbf{Positive Test Cases}. The set of positive test cases $T_\pi$ specifies the \emph{intended behavior} of a knowledge base (feature model). Positive test cases $t_i \in T_\pi$ are assumed to be existentially quantified, i.e., for each $t_i$ there should exist at least one configuration consistent with $t_i$. Such test cases can be \emph{derived} from already existing consistent complete or partial configurations (e.g., represented by a set of included and excluded features), from a set of analysis operations (e.g., dead features), or \emph{specified} by domain experts interested in  the correctness of the feature model. Without loss of generality, we restrict our working example to positive test cases specifying the intended behavior of a knowledge base (see Table \ref{tab:testcases}).

\begin{table}[ht]
\centering \caption{Example positive test cases $T_\pi=\{t_1,..,t_4\}$.}
\begin{tabular}{|c|c|c|c|c|c|c|c|c|}
\hline
  ID & Test Case (Constraint)                                 \tabularnewline
\hline
 $t_1$ & $nolicense = t$                            \tabularnewline
\hline
 $t_2$ & $license = t  \land statistics = f$      \tabularnewline
\hline
 $t_3$ & $payment = f$                             \tabularnewline
\hline
 $t_4$ & $singlechoice = f$                             \tabularnewline
\hline
\end{tabular} 
\label{tab:testcases} 
\end{table}

Avoiding \emph{nolicense} to be a \emph{dead feature} can be achieved by a positive test case $t_1: nolicense = t$ (\emph{nolicense} should be included at least one configuration). Similarly, an example of a partial \emph{survey} configuration could be: $t_2: license = t \land statistics = f$. Another test case could require the support of configurations with \emph{payment} being deactivated: $t_3: payment = f$. Finally, we introduce a test case that assures the existence of a configuration where \emph{singlechoice} is not included: $t_4: singlechoice = f$.

Note that test cases are not restricted to basic feature assignments but can also be implemented as general constraints, for example, we could specify $ABtesting \land license \rightarrow statistics$ as a test case. In the following, we will integrate  $\{t_1..t_4\}$ in our discussion of a diagnosis algorithm that supports the automated testing and debugging of feature models.

\textbf{Negative Test Cases}. Negative test cases $t_i \in T_\Theta$ can be regarded as all-quantified constraints which specify an \emph{unintended behavior} of a knowledge base. If $t_i$ is unexpectedly consistent with the knowledge base, it is integrated in negated form into the background knowledge (see Section \ref{automatedtestingdebugging}). 

\textbf{Generating Test Cases}. \emph{"Where do the test cases come from?"} is an important question to be answered for applying the presented concepts in industrial settings. First, positive test cases can be derived from already  \emph{completed and consistent feature model configurations}. Second, positive as well as negative test cases can be \emph{specified by domain experts}. Third, negative test cases can be derived from \emph{inconsistent configurations}, for example, configurations that have been identified as faulty by domain experts. Finally, test cases could also be \emph{directly generated from a feature model knowledge base}, for example, by using well-formedness criteria from feature model analysis operations \cite{BeSeRu2010} (e.g., to avoid \emph{dead features} $f_i$, a test case $f_i=t$ can be generated for each feature).  

\section{Automated Debugging with \textsc{DirectDebug}}\label{automatedtestingdebugging}

A diagnosis ($\Delta$) includes exactly those constraints responsible for the faulty behavior of a feature model. Intuitively, constraints in $\Delta$ have to be deleted or adapted to make the feature model consistent with  $T_\pi$ (see Definition 3). Thus, a diagnosis helps a feature model engineer to focus diagnosis search. To enable such a functionality, test cases are needed.

\textbf{Definition 3 (Diagnosis and Maximal Satisfiable Subset)}. Given a feature model with a set $CF$ of feature model constraints and a  set of positive test cases $T_\pi = \{t_1, t_2, .., t_l\}$. A diagnosis $\Delta=\{c_1, c_2, .., c_q\}$ is a  set of feature model constraints ($\Delta \subseteq CF$) such that $\forall t_i \in T_\pi: \{t_i\} \cup CF - \Delta$ is consistent. $\Delta$ is minimal \emph{iff} $\neg \exists \Delta' \subset \Delta$ such that $\forall t_i \in T_\pi: \{t_i\} \cup CF - \Delta'$ consistent. A complement of $\Delta$ (i.e., $CF - \Delta$) is denoted as \emph{Maximal Satisfiable Subset} (MSS $\Gamma$).

\begin{figure*}[ht]
			\centering
			\fbox{
				\includegraphics[width=0.95 \textwidth]{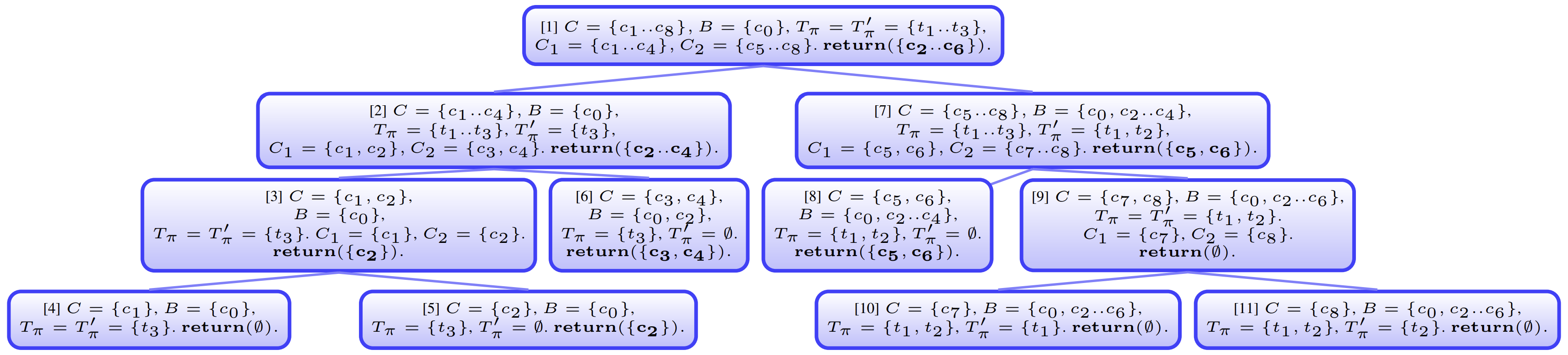}
			}
			\caption{\textsc{DirectDebug} execution trace for $C=\{c_1..c_8\}$, $B=\{c_0\}$, and $T_\pi=\{t_1 .. t_3\}$. Since $C \cup B \cup \{t_4\}$ is consistent, there is no need to further analyze $t_4$. \textsc{DirectDebug}  determines a maximal satisfiable subset MSS ($\Gamma= \{c_2..c_6\}$), the corresponding diagnosis is $\Delta = \{c_1,c_7,c_8\}$ (the MSS complement).}
			\label{fig:debugtree} \vspace{-0.35cm}
		\end{figure*}

\textbf{Example Diagnoses}. The minimal diagnoses that can be derived in our working example are depicted in Table \ref{tab:diagnoses}. There are two options for restoring the consistency between the feature model (Figure 1) and $\{t_1..t_4\}$: either delete/adapt the constraints in $\Delta_1$ or do this with the constraints of $\Delta_2$.  For example, $\Delta_1$ suggests to take a look at $c_1$ (reconsider the relationship between \emph{payment} and \emph{survey}) and $c_2$ (reconsider the relationship between \emph{ABtesting} and \emph{survey}).

\begin{table}[ht]
\centering  \caption{Example diagnoses $\Delta_1=\{c_1,c_2\}$ and $\Delta_2=\{c_1,c_7,c_8\}$.}
\begin{tabular}{|c|c|c|c|c|c|c|c|c|}
\hline
  Diagnosis & $c_1$ & $c_2$ & $c_3$ & $c_4$ & $c_5$ & $c_6$ & $c_7$ & $c_8$                               \tabularnewline
\hline
$\Delta_1$ & $\times$ & $\times$ & $-$ & $-$ & $-$ & $-$ & $-$ & $-$                               \tabularnewline
\hline
$\Delta_2$ & $\times$ & $-$ & $-$ & $-$ & $-$ & $-$ & $\times$ & $\times$                               \tabularnewline
\hline
\end{tabular} 
\label{tab:diagnoses} 
\end{table}

\textbf{Diagnosis Approach}. \textsc{DirectDebug} determines minimal diagnoses directly \cite{Felfernig2012}, i.e., without  predetermining conflicts. It extends direct diagnosis with test cases ($T_\pi$). \textsc{DirectDebug} is activated with the diagnosis candidates $C$ (constraints of $CF$ considered as potentially faulty) and the background knowledge $B$ (constraints of $CF$ assumed to be consistent and correct). The algorithm (Algorithm 1) determines an MSS $\Gamma$, the corresponding minimal diagnosis is $C-\Gamma$.

\begin{algorithm} \small
\caption{$\textsc{DirectDebug}(C=\{c_1..c_n\},B,T_\pi): \Gamma$}
\begin{algorithmic} \label{alg:qx}
\IF{\textsc{IsConsistent}($C \cup B,T_\pi,T_\pi'$)}
\STATE return($C$)
\ENDIF
\IF{$|C|=1$}
\STATE return($\emptyset$)
\ENDIF
\STATE $k=\lfloor\frac{n}{2}\rfloor$
\STATE $C_1 \leftarrow $ $c_1 ... c_{k}$; $C_2 \leftarrow $ $c_{k+1} ... c_{n}$;
\STATE $\Gamma_2 \leftarrow $ $\textsc{DirectDebug}(C_1,B,T_\pi')$; 
\STATE $\Gamma_1 \leftarrow $ $\textsc{DirectDebug}(C_2,B \cup \Gamma_2,T_\pi')$;
\STATE $return$($\Gamma_1 \cup \Gamma_2$) 
\end{algorithmic} 
\end{algorithm} 

The constraint $c_0$ should not be diagnosable, since empty feature models should not be allowed (this would be the case if $c_0$ is part of a diagnosis). We assume $c_0: survey = t$ to be part of the \emph{background knowledge} $B$ which consists of constraints assumed to be correct, i.e., $B$ entails those constraints which should not be regarded as diagnosis candidates. Furthermore, $B$ includes those \emph{negative test cases} $t_i \in T_\Theta$ \emph{in negated form} which are consistent with $C \cup B$. In our example, we assume $T_\Theta=\emptyset$ for simplicity. Before starting \textsc{DirectDebug}, all $t_i$ in $T_\pi= \{t_1 .. t_q\}$ and $T_\Theta= \{t_{q+1} .. t_z\}$ have to be checked for consistency with $C \cup B$. 

If at least one positive test case induces an inconsistency in $C \cup B$,  \textsc{DirectDebug} is activated. Please note that only those test cases in $T_\pi$ are forwarded to \textsc{DirectDebug} which are inconsistent with $C \cup B$. In our setting, the original set of positive test cases $T_\pi = \{t_1 .. t_4\}$ is reduced to $\{t_1 .. t_3\}$ since $t_4$ does not induce an inconsistency in $C \cup B$, i.e., there exists a configuration (solution) for $C \cup B \cup \{t_4\}$.
	
\textsc{DirectDebug} determines a maximal satisfiable subset MSS ($\Gamma$) (Definition 3) where $C \subseteq CF - \{c_0\}$ (consideration set), $B = CF - C \cup \{c_0\} \cup T^-$  (background knowledge), and $T^-$ represents a conjunction of negated negative test cases which are (unexpectedly) consistent with $C \cup B$. \textsc{DirectDebug} is activated with test cases $T_\pi$ that are inconsistent with $C \cup B$. $C \subseteq CF - \{c_0\}$ can be used to focus diagnosis search on specific parts of a feature model. If $CF$ should be diagnosed as a whole, $C = CF - \{c_0\}$ and $B = \{c_0\} \cup T^-$.

\textbf{\textsc{DirectDebug} Consistency Checks}. \textsc{IsConsistent} checks whether the constraints in $C \cup B$ are consistent with the  test cases in $T_\pi$. Obviously, test cases have to be checked individually, i.e., each activation of \textsc{IsConsistent} results in $|T_\pi|$ constraint solver activations (consistency checks).  \textsc{IsConsistent} returns $t$ ($true$) if every test case in $T_\pi$ is consistent with $C \cup B$, otherwise $f$ ($false$). Only test cases inducing an inconsistency with $C \cup B$ are stored (returned) in $T_\pi'$ (the remaining inconsistent positive test cases). 

\textbf{\textsc{DirectDebug} Execution}. An execution trace of  \textsc{DirectDebug} is shown in Figure \ref{fig:debugtree}. \textsc{DirectDebug} follows a \emph{divide \& conquer} approach. In each incarnation, it is analyzed which  positive test cases remain inconsistent with $C \cup B$. If just one constraint $c_i$ remains in the consideration set $C$ ($|C|=1$) and there still exists at least one test case $t_j$ with inconsistent($\{t_j\} \cup C \cup B$), then $c_i$ is considered a part of a diagnosis $\Delta$. If $C \cup B$ is consistent with the test cases in $T_\pi$, $C$ is returned since no diagnosis elements can be found in $C$. 

\textbf{Diagnosis Determination}. \textsc{DirectDebug} returns a \emph{maximal satisfiable subset} ($\Gamma = \{c_2..c_6\}$) (see Figure \ref{fig:debugtree}). To determine a minimal diagnosis $\Delta$, we have to determine the MSS-complement, i.e., $C - \Gamma$ which is $\{c_1,c_7,c_8\}$.

 \section{Performance Analysis}\label{evaluation} 

To evaluate the performance of \textsc{DirectDebug}, we synthesized test feature models (see Table \ref{tab:performance}). For model synthesis, we applied the \textsc{Betty} generator \cite{betty} using the parameters \emph{\#test positive cases} ($|T_\pi|$, with a 30\% share of inconsistency-inducing test cases) and \emph{\#constraints in} $CF$ ($|CF|$, where \#variables = $\frac{|CF|}{2}$) with $C = CF-\{c_0\}$. Since each test case check needs a constraint solver call\footnote{For evaluation purposes, we used choco-solver.org.}, runtimes  increase with an increasing number of test cases and constraints. Each $|T_\pi|$ $\times$ $|CF|$ entry in Table \ref{tab:performance} represents the average \textsc{DirectDebug} (diagnosis) computing time after 3 repetitions.

\begin{table}[ht]
\centering 
\caption{Runtimes (in $msec$) of \textsc{DirectDebug} with different constraint set ($CF$) and test set ($T_{\pi}$) cardinalities ($\left|T_{\Theta}\right|=0$) evaluated on an Intel Core i7 (6 cores) 2.60GHz  with 16 GB of RAM.}

\vspace{-0.25cm}

\begin{tabular}{|l|c|c|c|c|c|c|c|c|}
\hline
 & \multicolumn{6}{c|}{$|CF|$} \tabularnewline
\cline{2-7}
$|T_\pi|$ & 10     & 20 & 50 & 100 & 500 & 1000   \tabularnewline
\hline
5             &  0.2   &  0.4  &  1.5  &  3.6   &  31.8   & 134.7
\tabularnewline
\hline
10             &   0.3  &  0.6  &  2.1 &  6.0   &   43.8  & 200.5
\tabularnewline
\hline
25             &  0.7   &  1.7  &  5.2  &  15.3   &  127.9   & 441.9
\tabularnewline
\hline
50             &   1.3 &  2.9  &  9.3  &   27.3  &   275.8  & 807.2
\tabularnewline
\hline
100             &  2.8   &  5.3  &  16.8  &   45.5  &  500.6   & 1463.7
\tabularnewline
\hline
250             &   8.7  &  15.1  &  37.9  &  105.1   &  1297.6   & 3290.1
\tabularnewline
\hline
500             &  27.0   &  27.7  &  68.1 &   182.1  &   2526.7  & 6429.0
\tabularnewline
\hline
\end{tabular} 
\label{tab:performance} 
\end{table}

\section{Conclusions}\label{conclusions}
We have introduced an approach to the automated testing and debugging of feature models. Test cases can be used to induce conflicts in the knowledge base (representing a feature model). Using \emph{direct diagnosis}, we show how consistency can be restored using a minimal diagnosis that includes all constraints responsible for the inconsistency. With this, we can pro-actively support feature model designers and can expect significant time savings in feature model development and evolution. Major issues for future work include the development of techniques for the automated generation of test cases taking into account coverage metrics that help to focus search on the most relevant parts of a knowledge base. Furthermore, we intend to include information from feature model quality metrics to better predict the most relevant diagnoses. Finally, we will continue our evaluations with real-world feature models.

\section{Data
Availability}\label{sec:data} The test dataset used for evaluation purposes can be found here: \emph{https://github.com/AIG-ist-tugraz/DirectDebug}.

\section*{Acknowledgment}  This work has been partially funded by the Horizon 2020 project \textsc{OpenReq} (732463), the Austrian Research Promotion Agency \textsc{ParXCel} project (880657), and the EU FEDER program MINECO project \textsc{Ophelia} (RTI2018-101204-B-C22).

\bibliographystyle{IEEEtran}
\bibliography{IEEEabrv,bibliography}

\end{document}